\documentclass{article}
\usepackage{amssymb}
\usepackage{epsfig}
\usepackage{amsmath}

\newcommand{\Z}{{\mathbb Z}}
\newcommand{\I}{{\mathbb I}}
\newcommand{\R}{{\mathbb R}}
\def\mathcenter#1{%
  \vcenter{\hbox{#1}}%
}

\newcommand{\mfig}[2][]{
        \mathcenter{\includegraphics[#1]{#2}}
}
\input{tcilatex}

\begin{document}

\title{\textbf{General relativity, Lauricella's hypergeometric function
 $F_D$ and the theory of braids}}
\author{G. V. Kraniotis \footnote{kranioti@mppmu.mpg.de} \\
%EndAName
Max Planck Institut f$\rm{\ddot u}$r Physik,\\
F$\rm \ddot{o}$hringer Ring 6,\\
D-80805 M$\rm{\ddot u}$nchen, Germany 
\footnote{MPP-2007-133, September 2007} \\
}
\maketitle

\begin{abstract}
The exact (closed form) solutions of the equations of motion in the theory 
of general relativity that describe motion of test particle and photon in 
Kerr and Kerr-(anti) de Sitter spacetimes all involve the multivariable 
hypergeometric function of Lauricella $F_D$: 
Kraniotis [Class. Quantum Grav. {\bf 21} 2004, 4743;
Class. Quantum Grav. {\bf 22} 2005, 4391; Class. Quantum Grav. {\bf 24} 2007, 1775]. The domain of variables ${\cal D}_n$ of the corresponding function depends on the first integrals of motion 
associated with the isometries of the Kerr-(anti) de Sitter metric and 
Carter's constant $Q$ as well as on the cosmological constant $\Lambda$ and 
the Kerr (rotation) parameter. In this work we discuss the topological properties of the domain ${\cal D}_n$ and in particular its fundamental connection with 
the theory of braids. An intrinsic relationship of general relativity  with 
the pure braids is established.

\end{abstract}

\bigskip

\bigskip \newpage

\section{ Introduction}

\subsection{\protect\bigskip Motivation}

The exact (closed form) solutions 
of the equations of motion of the theory 
of general relativity (GTR)  that describe orbits  of test particle and 
photon in 
Kerr and Kerr-(anti) de Sitter spacetimes
have yielded the following result \cite{KraniotisKerr},\cite{KraniotisLight},
\cite{KraniotisSstars}: 

All the physical amplitudes (measurable quantities) related 
to test particle orbits such as periapsis and 
gravitomagnetic (Lense-Thirring) precessions, orbital periods 
as well as the bending of light  
by a rotating central mass (rotating black hole or rotating star), 
the gravitomagnetic precessions and orbital periods of spherical 
photon orbits in Kerr spacetime with a cosmological constant have been 
elegantly expressed in terms of  Lauricella's multivariable 
hypergeometric function $F_D(\alpha,\beta_1,\beta_2,
\cdots,\beta_m,\gamma;z_1,z_2,\cdots,z_m)$.
The domain of variables (moduli) of Lauricella's function $F_D$ \cite{APPELL,
LAURICELLA}
\begin{equation}
{\cal D}_n={(z_1,z_2,\cdots,z_n);z_i \not= 0,1,(1\leq i\leq n),z_j\not =
z_k (1\leq j<k\leq n)}
\end{equation}
is related in the theory of General Relativity through the exact 
solutions of the geodesics system in Kerr-(anti) de Sitter spacetime 
to the first integrals of motion, as well as 
to the cosmological constant $\Lambda$ and the rotation (Kerr) parameter \cite{
KraniotisKerr},\cite{KraniotisLight},\cite{KraniotisSstars}. 
The generalised hypergeometric function of Appell-Lauricella $F_D$ is a 
very important function in Mathematical Analysis  and as we shall see in 
this work it possesses very interesting topological properties. 
This then can lead to a fundamental relationship of General Relativity 
with topology. The establishment of such a relationship is the main 
theme and objective of this work.

Indeed in pure Mathematics the domain of variables of 
$F_D$ has been studied \cite{Terada}; a main result of this investigation  
was the very interesting topological properties of the domain 
${\cal D}_n$.  Essentially, the fundamental group of 
the domain under discussion, $\pi_1({\cal D}_n,a)$, crudely speaking is the 
pure (or coloured) symmetry braid group \cite{Terada}. 
Thus the results of  \cite{
KraniotisKerr},\cite{KraniotisLight},\cite{KraniotisSstars} 
combined with the previous result constitute  a profound and intrinsic relation of the theory of 
General Relativity with the field of algebraic topology and in particular 
with the theory of braids and links \footnote{As we shall see in the 
main body of the paper the closure of braids are knots and links. In 
particular, the closure of a coloured braid is a link. }. The {\em first}, 
as a matter of fact, {\em direct connection} of a theory of Physics  
with the theory of braids \footnote{At this point we must mention 
that nice accounts of previous efforts and results connecting certain models 
in physics with topological invariants can be found in \cite{Atiyah,Prasolov}.}.

The material of this paper is organized as follows:  In section 
\ref{BraidsBstuff} we present some basic results of braid theory (
which are useful in understanding the main result); 
Namely the presentation theory of the braid and pure braid group as 
well as their representation theory through the Burau and Gassner 
matrices respectively (subsections \ref{Buraur} and \ref{Gassner}).
Having discussed the closure operation and the Markov moves in section
\ref{MarkovClosure}, the topology invariants for knots and links 
associated with the discovery of Jones, Bracket and HOMFLY polynomials 
are briefly discussed and reviewed in \ref{KaufJones}.
In section \ref{ThemelioOmada1} we discuss the presentation of 
$\pi_1({\cal D}_n,a)$ and the hypergeometric representation of the 
pure or coloured group that it defines, an approach 
developed in \cite{Terada}. Combining with the results in \cite{
KraniotisKerr},\cite{KraniotisLight},\cite{KraniotisSstars} 
the establishment of the 
fundamental connection of the theory of General Relativity with 
the coloured braids via the hypergeometric function of Lauricella 
$F_D$ is then achieved.
Finally, section \ref{Sympera} is used for our conclusions.

\section{The theory of braids and their symmetry group}
\label{BraidsBstuff}

Braids (Z$\rm \ddot{o}$pfe) are very beautiful and profound mathematical entities. They have been 
constructed by the German mathematician  Emil Artin \cite{Artin},\cite{Emil} 
first as an application 
to textile industry evolving into a central theme of topology 
where they currently serve as the fundamental theory of knots and links.

\begin{figure}[h]
\begin{center}
\includegraphics[width=3.7in]{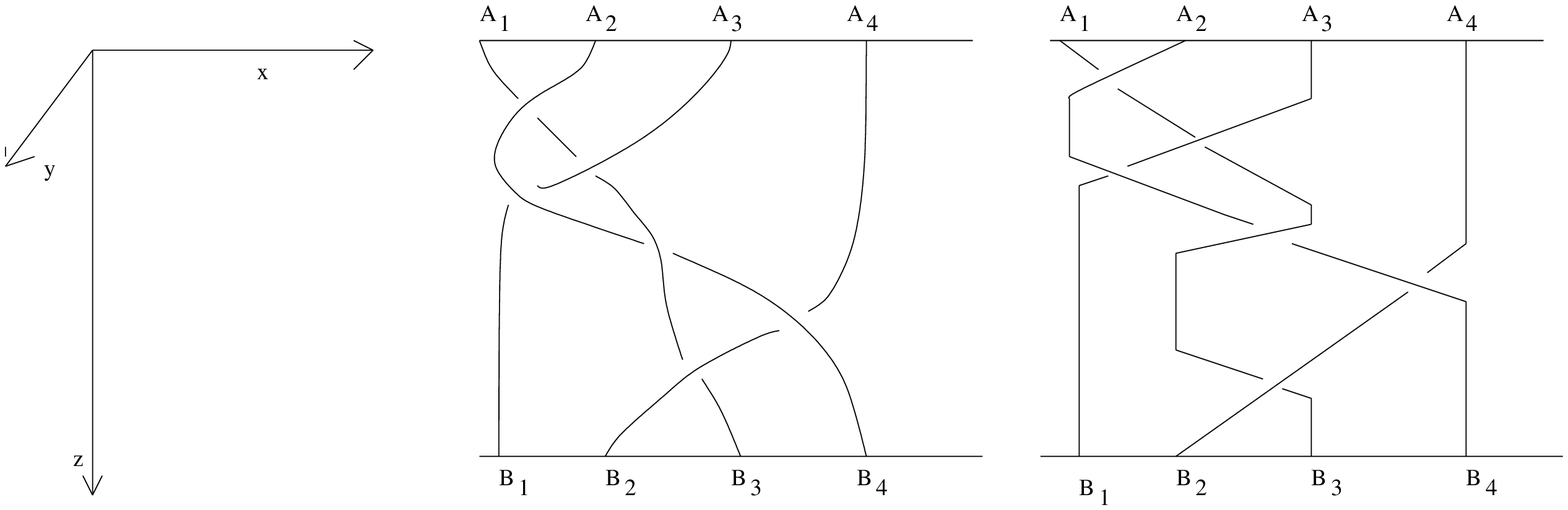}
\end{center}
\caption{Examples of braids.}
\label{Braids}
\end{figure}

In the space $\R^3$, consider the points $A_i=(i,0,0)$ and $B_i=(i,0,1)$, 
where $i=1,2,\ldots,n$. A polygonal line joining one of the points $A_i$ 
with one of the points $B_j$ is  called ascending if in the motion of a 
point from $A_i$ to $B_j$ along the line its z-coordinate increases 
monotonically.
A braid in $n$ strands (or strings) is defined as  a set of pairwise nonintersecting ascending polygonal lines (the strands) joining the points 
$A_1,\ldots,A_n$ to the points $B_1,\ldots,B_n$ (in any order).
One can also consider braids whose strands are ascending smooth lines (rather 
than polygonal ones); then it is natural to define equivalence as isotopy, 
i.e., as a smooth deformation in the class of braids. Examples of braids are 
given in fig.\ref{Braids}.

Braids form a group and we now discuss its properties.

\begin{figure}[h]
\begin{center}
\includegraphics[width=1.7in]{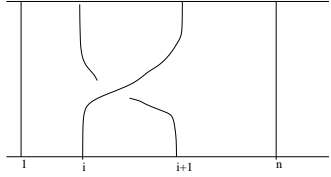}
\end{center}
\caption{The generator $\sigma_i$.}
\label{genitorasB}
\end{figure}

We denote by $\sigma_i$ the braid which joins $i$ to $i+1$ by a path 
passing under the path joining $i+1$ to $i$ (see Figure \ref{genitorasB}).

The braid group $B_n$ is generated by the elements $\sigma_1,\sigma_2,\cdots\sigma_{n-1}$ with the following presentation \cite{Artin}:

\begin{eqnarray}
\rm {Generators}: && \sigma_1,\sigma_2,\cdots\sigma_{n-1} \nonumber \\
\rm {Relations}: && \sigma_i\sigma_{i+1}\sigma_i=\sigma_{i+1}\sigma_i\sigma_{i+1},\;\;\;(i=1,\cdots,n-2) \nonumber \\
&& \sigma_i\sigma_j=\sigma_j\sigma_i \;\;\;{\rm for}\;\;\; |i-j|>1 \nonumber \\
\end{eqnarray}

The second relation is sometimes called {\it far commutativity}, because 
it says the generators commute pairwise when they are sufficiently far from 
each other, i.e., when their indices differ by two or more.

\subsection{Relationship between braids and links and the closure operation}
\label{MarkovClosure}

In this section we discuss the relationship betweeen braids and links 
arising from the closure operation, which assigns a knot, link to each 
braid in a natural way.

%\begin{figure}
%\centering
%\mfig[scale=0.3]{closure.eps}
%\caption{Closure of a braid}
%\end{figure}

\begin{figure}
\begin{center}
\includegraphics[width=1.7in]{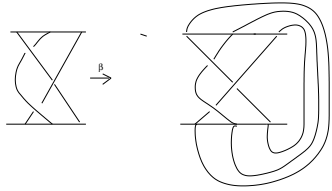}
\end{center}
\caption{Closure of a braid}
\label{Closure}
\end{figure}

There is a {\it canonical epimorphism} $\varphi:B_n \rightarrow S_n$ of the 
braid group onto the permutation group.
In terms of relations, the group $S_n$ is obtained from $B_n$ by adding 
the relation $\sigma_i^2=1$.
A link or knot $\beta(b)$  is obtained by closing the braid $b$, i.e., 
tying the top end of each string (strand) to the same position at the 
bottom of the braid as shown in fig.\ref{Closure}. 
The closure $\beta(b)$ of the braid $b$ is a knot if the permutation $\varphi(b)$ associated to the braid generates the cyclic subgroup of order $n$, 
$\Z/n\Z$, in the permutation group $S_n$ \cite{Prasolov}.

Next we discuss an important theorem due to A. Markov which answers the 
question of when different braids can have isotopic closures, i.e. represent the same 
knot, link. The {\em first Markov move} replaces $b\in B_n$ by 
$aba^{-1}$ for $a\in B_n$. The {\em second Markov move} is the replacement 
$b\leftrightarrow b \sigma_n^{\pm 1}$ for $b\in B_n$ (note that $\sigma_n \not \in B_n$, so the notation $b\sigma_n$ makes sense algebraically only if 
we identify $b$ with its image under the natural inclusion $B_n\hookrightarrow
B_{n+1}$) Then the theorem asserts that the closures of two braids are isotopic if and only if one braid can be taken to another by a finite sequence of 
Markov moves. A proof of this theorem can be found in the book of J. S. Birman 
\cite{Joan}.

Despite the difficulty of applying Markov's theorem for studying knots via braids,  braids had first suggested themselves as a useful tool for investigating 
further their relationship with links via the closure operation, after the discovery by Werner Burau of a matrix representation of $B_n$. This representation 
is the subject of the following section

\subsubsection{The Burau representation}  
\label{Buraur}

If $x$ is a non-zero complex number let $M_i$, for $1\leq i \leq n-1$, be 
the $n\times n$ matrix

%$$M_i=\left(\begin{array}{ccccccc}
%1&&&&&&0 \\
%&\ddots&&&&&\\
%&&1-x &x&&& \\
%%&&1&0&&& \\
%%&&&&1&& \\
%%&&&&&\ddots& \\
%%0&&&&&&1
%%\end{array}\right)$$

$$\bordermatrix{& & & & & & & \cr
&1&&&&&&0 \cr
& &\ddots&&&&& \cr
i &&&1-x &x&&& \cr
i+1&&&1&0&&& \cr
&&&&&1&& \cr
&&&&&&\ddots& \cr
&0&&&&&&1 }=M_i$$

where $1-x$ is the $i-i$ entry. One may easily check that 
$M_i M_{i+1}M_i=M_{i+1}M_i M_{i+1}$ and 
$M_i M_j=M_jM_i$ if $|i-j|\geq 2$. Thus sending 
$\sigma_i$ to $M_i$ defines the (non-reduced) Burau representation 
of $B_n$ \cite{WBurau}.
Burau recognized that his representation  was related to closed braids.
More specifically if $\alpha\in B_n$ and $\psi$ is the reduced Burau 
representation then ${\rm det} (1-\psi(\alpha))$ is $(1+x+\ldots+x^{n-1})$ 
times the Alexander polynomial of the link $\hat{\alpha}$ \cite{Alexander}.

\subsection{The coloured braid group symmetry}
\label{Gassner}

The kerner of the epimorphism $\varphi$ defines the {\em coloured braid 
symmetry group}  or {\em pure braid group} $P_n$ \footnote{As a matter of 
fact, the following short exact sequence is valid: $1 \rightarrow 
P_n\overset{\rho}{ \rightarrow} B_n \overset{\varphi}{\rightarrow} S_n \rightarrow 1$.}

\begin{equation}
P_n={\rm Ker}{\varphi}
\end{equation}

We first define generators for $P_n$.

For any $i<j$, set  $A_{ij}=A_{ji}=\sigma_{j-1}\sigma_{j-2}\cdots 
\sigma_{i+1}\sigma_i^2 \sigma_{i+1}^{-1}\cdots \sigma_{j-2}^{-1}\sigma_{j-1}^{-1}$.

\begin{figure}[h]
\begin{center}
\includegraphics[width=1.7in]{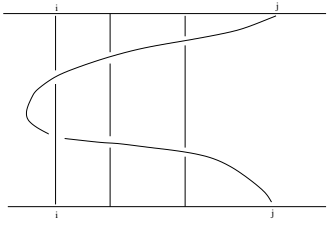}
\end{center}
\caption{The generator $A_{ij}$ for the pure symmetry braid group.}
\label{ErzeugendP}
\end{figure}

The generator $A_{ij}$ is depicted in Figure \ref{ErzeugendP}.

Artin \cite{Emil} gives the following presentation for $P_n$ \footnote{One way 
to obtain a presentation for $P_n$ is using the Schreier-Reidemeister 
method. The group $P_n$ is of index $n!$ in $B_n$. One may choose 
as coset representatives for $P_n$ in $B_n$ any set of $n!$ words in the 
generators of $B_n$ whose images under $\varphi$ range over all of $S_n$.
When these coset representatives form a Schreier set (i.e. any initial 
segment of a coset representative is again a coset representative ) 
one can apply the Schreier-Reidemeister 
method \cite{Otto}.}

\begin{eqnarray}
& &{\rm Generators}:  A_{ij};1\leq i<j<n \nonumber \\
& &{\rm Relations}: \nonumber \\
& & 1) \;\;\;A_{rs}^{\epsilon}A_{ik}A_{rs}^{-\epsilon}=
A_{ik} \nonumber \\
& & {\rm if\; all\; indices\; are\; different} \nonumber \\
& & {\rm and\; if\; the\; pairs}\;  r,s\; {\rm}and\; i,k\; {\rm do\; not\; 
separate\; each\; other} \nonumber \\
& & 2) \;\;\; A_{rs}^{\epsilon}A_{ir}A_{rs}^{-\epsilon}= A_{is}^{-\epsilon}A_{ir}A_{is}^{\epsilon} \nonumber \\
& & 3) \;\;\;A_{rs}^{\epsilon}A_{is}A_{rs}^{-\epsilon}=
A_{is}^{-\epsilon}A_{ir}^{-\epsilon}A_{is}A_{ir}^{\epsilon}A_{is}^{\epsilon} 
\nonumber \\
& &{\rm if\; finally\; the\; subscripts\; are\; all\; different\;and\;}\nonumber \\
& &{\rm the\; pairs\;} r,s\; {\rm and}\; i,k\; {\rm separate\; each\; other\; we \;get} \nonumber \\
& &4)\;\;\; A_{rs}^{\epsilon}A_{ik}A_{rs}^{-\epsilon}=A_{is}^{-\epsilon}A_{ir}^{-\epsilon}
A_{is}^{\epsilon}A_{ir}^{\epsilon}. A_{ik}. A_{ir}^{-\epsilon}
A_{is}^{-\epsilon}A_{ir}^{\epsilon}A_{is}^{\epsilon} \nonumber \\
\end{eqnarray}
and $\epsilon=\pm 1$.

Every pure braid can be combed i.e. it can be represented in terms of the 
generators $A_{ij}$

\begin{eqnarray}
\sigma_1\sigma_2\sigma_2\sigma_1 &=& A_{12}A_{13} \nonumber \\
\, \mfig[scale=0.3]{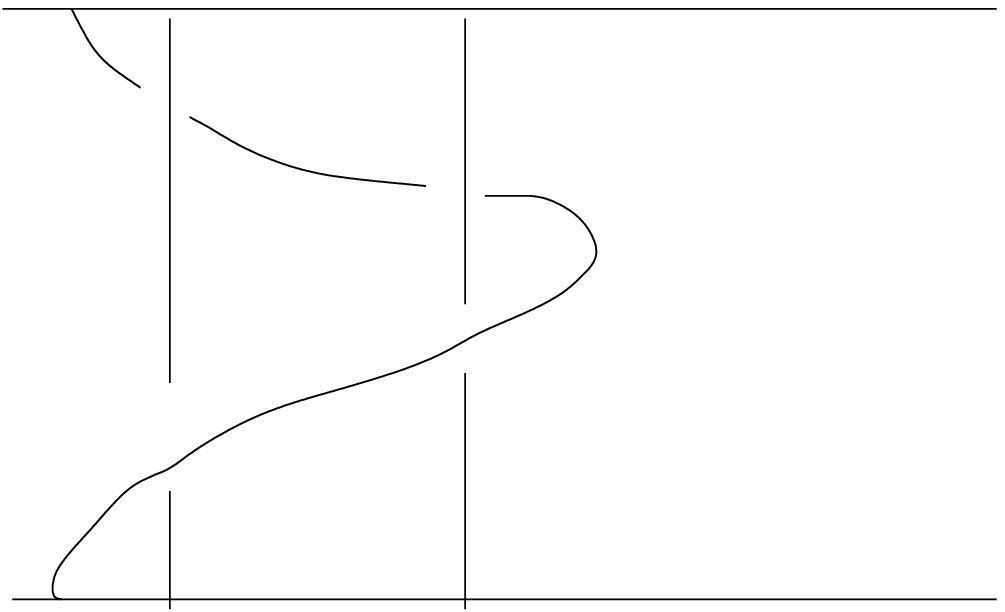} \,&=&
\, \mfig[scale=0.22]{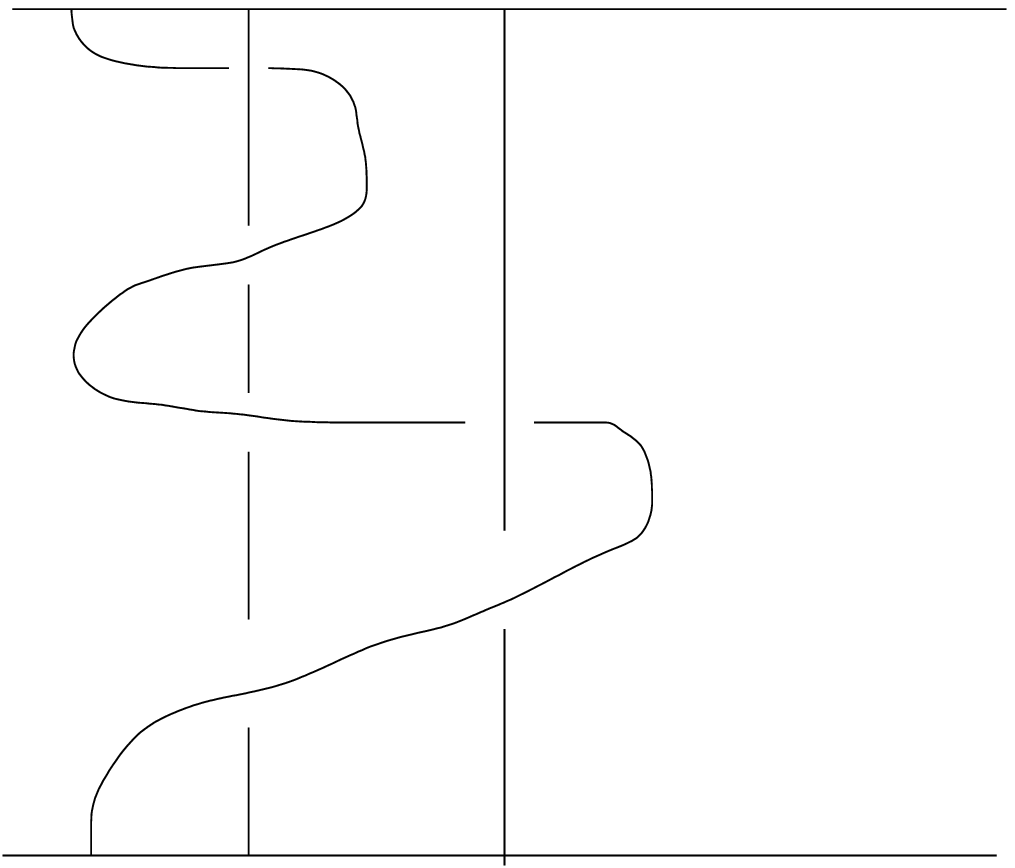} \, \nonumber \\
\end{eqnarray}

The element $(A_{12})(A_{13}A_{23})\cdots(A_{1n}A_{2n}\cdots A_{n-1 n})\in$ 
centre of $P_n$.

Let us give an example for $n=4$, the symmetry group $P_4$ has presentation

\begin{eqnarray}
& &{\rm Generators}:  A_{12},A_{23},A_{34},A_{13},A_{24},A_{14} \nonumber \\
& &{\rm Relations}: \nonumber \\
& & A_{12}A_{34}=A_{34}A_{12}  \nonumber \\
& & A_{14}A_{23}=A_{23}A_{14} \nonumber \\
& & A_{24}A_{13}A_{24}^{-1}=A_{14}^{-1}A_{12}^{-1}A_{14}A_{12}A_{13}A_{12}^{-1}A_{14}^{-1}A_{12}A_{14} \nonumber \\
& & A_{23}A_{12}A_{23}^{-1}=A_{13}^{-1}A_{12}A_{13} \nonumber \\
& & A_{24}A_{12}A_{24}^{-1}=A_{14}^{-1}A_{12}A_{14} \nonumber \\
& & A_{34}A_{13}A_{34}^{-1}=A_{14}^{-1}A_{13}A_{14} \nonumber \\
& & A_{34}A_{23}A_{34}^{-1}=A_{24}^{-1}A_{23}A_{24} \nonumber \\
& & A_{23}A_{13}A_{23}^{-1}=A_{13}^{-1}A_{12}^{-1}A_{13}A_{12}A_{13} \nonumber \\
& & A_{24}A_{14}A_{24}^{-1}=A_{14}^{-1}A_{12}^{-1}A_{14}A_{12}A_{14} \nonumber \\
& & A_{34}A_{14}A_{34}^{-1}=A_{14}^{-1}A_{13}^{-1}A_{14}A_{13}A_{14} \nonumber \\
& & A_{34}A_{24}A_{34}^{-1}=A_{24}^{-1}A_{23}^{-1}A_{24}A_{23}A_{24} \nonumber \\
\label{Kath4}
\end{eqnarray}

It can easily be checked that the following matrices constitute a representation of the pure braid group $P_4$, i.e. they satisfy the relations (\ref{Kath4}).

$$A_{12}=\left(\begin{array}{cccc}
1-x_1+x_1 x_2&(1-x_1)x_1&0&0 \\
1-x_2&x_1&0&0\\
0&0&1 &0 \\
0&0&0&1 \\
\end{array}\right)$$

$$A_{13}=\left(\begin{array}{cccc}
1-x_1+x_1 x_3&0&(1-x_1)x_1&0 \\
(1-x_2)(1-x_3)&1&(-1+x_1)(1-x_2)&0\\
1-x_3&0&x_1 &0 \\
0&0&0&1 \\
\end{array}\right)$$

$$A_{14}=\left(\begin{array}{cccc}
1-x_1+x_1 x_4&0&0&(1-x_1)x_1 \\
(1-x_2)(1-x_4)&1&0&(-1+x_1)(1-x_2)\\
(1-x_3)(1-x_4)&0&1 &(-1+x_1)(1-x_3) \\
1-x_4&0&0&x_1 \\
\end{array}\right)$$

$$A_{23}=\left(\begin{array}{cccc}
1&0&0&0 \\
0&1+x_2(-1+x_3)&(1-x_2)x_2&0\\
0&1-x_3&x_2 &0 \\
0&0&0&1 \\
\end{array}\right)$$

$$A_{24}=\left(\begin{array}{cccc}
1&0&0&0 \\
0&1-x_2+x_2x_4&0&(1-x_2)x_2\\
0&(1-x_3)(1-x_4)&1&(-1+x_2)(1-x_3) \\
0&1-x_4&0&x_2 \\
\end{array}\right)$$

$$A_{34}=\left(\begin{array}{cccc}
1&0&0&0 \\
0&1&0&0\\
0&0&1-x_3+x_3 x_4&(1-x_3)x_3 \\
0&0&1-x_4&x_3 \\
\end{array}\right)$$

This matrix representation is the famous representation 
discovered by Betty Jane Gassner in 1961 
\cite{BettyJane}, generalizing the Burau representation.

The element $A_{12}A_{13}A_{23}A_{14}A_{24}A_{34}$ is represented by the 
matrix

$$\left(\begin{array}{cccc}
1+x_1(-1+x_2x_3x_4) & -(-1+x_1)x_1 &-(-1+x_1) x_1x_2 & -(-1+x_1) x_1x_2x_3 \\
1-x_2 & x_1 (1+x_2(-1+x_3 x_4)) & -x_1 (-1+x_2)x_2 &-x_1(-1+x_2)x_2 x_3 \\
1-x_3 & x_1-x_1 x_3 & x_1 x_2(1+x_3 (-1+x_4)) & -x_1 x_2 (-1+x_3)x_3 \\
1-x_4 & x_1-x_1 x_4 & -x_1 x_2(-1+x_4) & x_1 x_2 x_3 \\
\end{array}\right)$$

More generally, denoting by $A_{rs},1\leq r<s\leq n$, the generators of 
$P_n$, the (unreduced) Gassner representation is the homomorphism 
$G_n:P_n\rightarrow GL_n(\Z[x_1^{\pm},\cdots,x_n^{\pm}])$ given by the 
formula \footnote{The faithfulness of $G_n$ for $n \geq 4$ is an important 
issue in braid theory.}
 
$$G_n(A_{rs})=\left(\begin{array}{ccccc}
\I_{r-1} & 0&0&0&0 \\
0& 1-x_r+x_rx_s & 0 & x_r(1-x_r) &0 \\
0 & \vec{u} & \I_{s-r-1} & \vec{v} &0 \\
0 & 1-x_s & 0 & x_r & 0 \\
0 & 0 &0 &0 & \I_{n-s} \\
\end{array}\right)$$

where 
\begin{equation}
\vec{u}=((1-x_{r+1})(1-x_s)\cdots(1-x_{s-1})(1-x_s))^{\top}
\end{equation}
and
\begin{equation}
\vec{v}=((1-x_{r+1})(x_r-1)\cdots(1-x_{s-1})(x_r-1))^{\top}
\end{equation}
and $\I_k$ denotes the $k\times k$ identity matrix.

\section{The Bracket polynomial}
\label{KaufJones}

Jones introduced his polynomial invariant for tame oriented links via 
certain representations of the braid group \cite{VFRJones} \footnote{Different from the Burau representation in section \ref{Buraur}.}, exploiting the similarity 
of the Ocneanu trace in Hecke algebras with the Markov moves, 
first pointed out to him by Joan S Birman.
There are two ways to introduce the Jones \cite{VFRJones} and HOMFLY 
(or LYMPHTOFU) polynomials \cite{FYHLMO}. First through braids (each knot and link expressed as a word in the generators 
of the braid group) \cite{VFRJones}, and second through the bracket polynomial due to L Kauffman
\cite{Kauffman}. We briefly discuss both approaches beginning with the definition and properties of the bracket polynomial \cite{Kauffman}.

\begin{figure}[h]
\begin{center}
\includegraphics[width=2.9in]{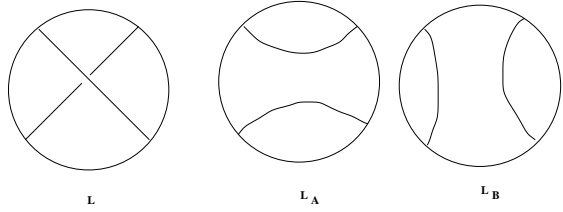}
\end{center}
\caption{Eliminating a crossing point.}
\label{Elimination}
\end{figure}

To each nonoriented link diagram $L$ a polynomial in the variables 
$a,b,c$ is assigned, denoted by $<L>$ which satisfies the following 
defining relations

\begin{equation}
\left < \, \mfig[scale=0.3]{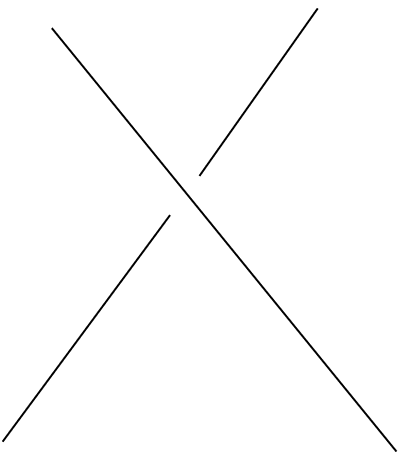} \, \right >
=a \left< \, \mfig[scale=0.3]{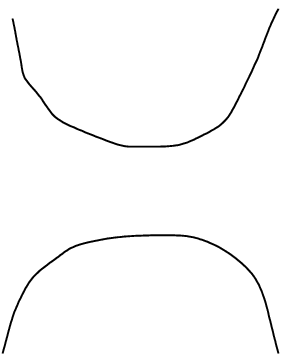} \, \right>+
b\left<\, \mfig[scale=0.3]{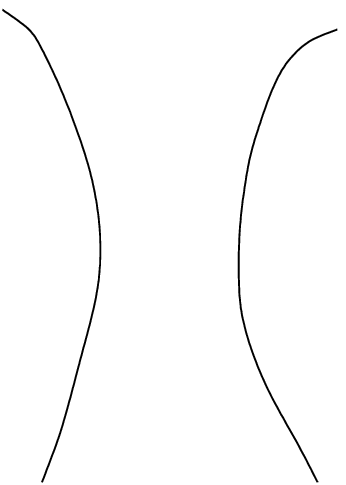} \, \right>
\label{KaufB}
\end{equation}

\begin{equation}
<L\sqcup0>=c<L>
\end{equation}

\begin{equation}
<0>=1
\end{equation}
Here the little pictures in (\ref{KaufB}) denote three link diagrams 
$L,L_A,L_B$ which are identical outside a small disk and are as 
shown in the picture \ref{Elimination} inside it.

\begin{figure}[h]
\begin{center}
\includegraphics[width=1.1in]{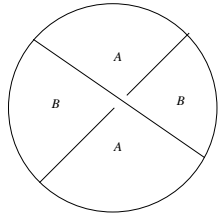}
\end{center}
\caption{$A$ and $B$ regions near a crossing point.}
\label{Regional}
\end{figure}

In this notation (\ref{KaufB}) may be rewritten as $<L>=a<L_A>+b<L_B>$.
The arcs inside the small disks of the diagrams $L_A$ and $L_B$ are chosen 
in the regions $A$ and $B$ defined in Figure \ref{Regional}

It turns out that the bracket polynomial is invariant under two of the 
Reidemeister moves which imposes the constraints $b=a^{-1},c=-a^2-b^2$.

For instance for the knot $4_1$ the bracket polynomial is calculated as 
follows
\begin{equation}
\left < \, \mfig[scale=0.3]{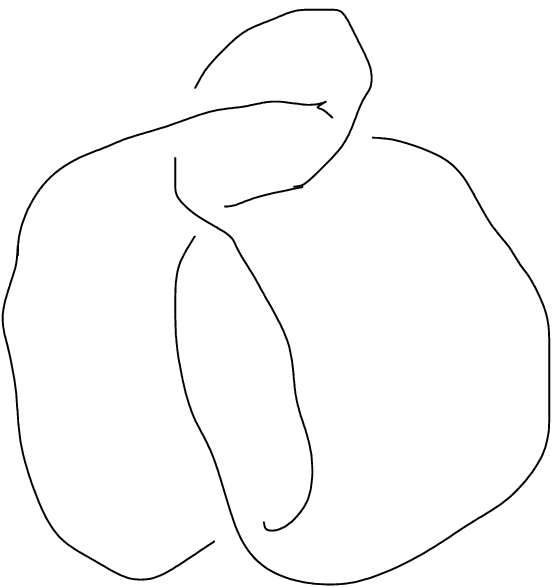} \, \right >=1+a^{-8}+a^8-a^4-a^{-4}
\label{kn41}
\end{equation}
while for the link below we have the result
\begin{equation}
\left < \, \mfig[scale=0.3]{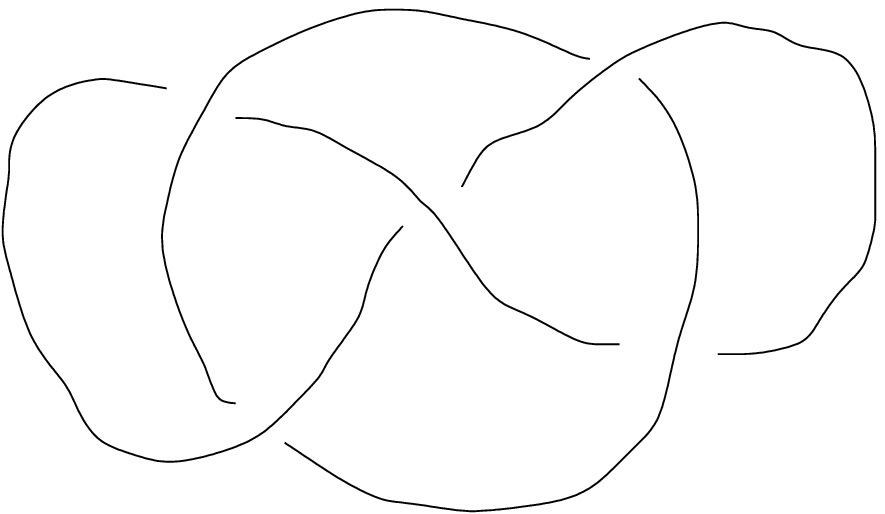} \, \right >=-a^{11}+2 a^7-a^3+2 a^{-1}-
a^{-5}+a^{-9}
\end{equation}

\begin{figure}[h]
\begin{center}
\includegraphics[width=1.9in]{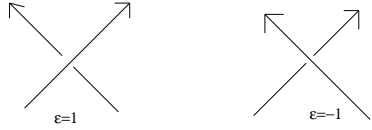}
\end{center}
\caption{Positive and negative crossing points.}
\label{Crossings}
\end{figure}

For the Kauffman polynomial one needs to consider {\em oriented} links, i.e. 
we assume that each component is supplied with an orientation (shown by 
arrows in the figures).  The {\em writhe number} is defined as follows 
\begin{equation}
\omega(L):=\sum_i \epsilon_i
\end{equation}
where the sum is taken over all crossing points and the numbers $\epsilon_i$ 
are equal to $\pm 1$ depending on the sign of the $i$th crossing point, which 
is defined in figure \ref{Crossings}.

Then the Kauffman polynomial $X(L)$ on any oriented link diagram $L$ is 
defined by \cite{Kauffman}
\begin{equation}
X(L):=(-a)^{-3 \omega(L)}<|L|>
\end{equation}
where the nonoriented diagram $|L|$ is obtained from $L$ by forgetting the 
orientation of all components.
Now for the knot below the Kauffman polynomial is calculated to be
\begin{eqnarray}
X\left(\, \mfig[scale=0.3]{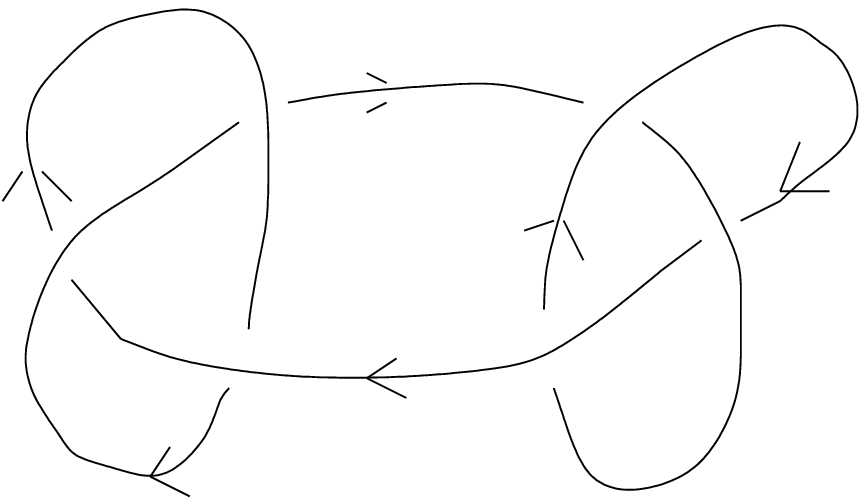} \,\right)&=&\left < \, \mfig[scale=0.3]{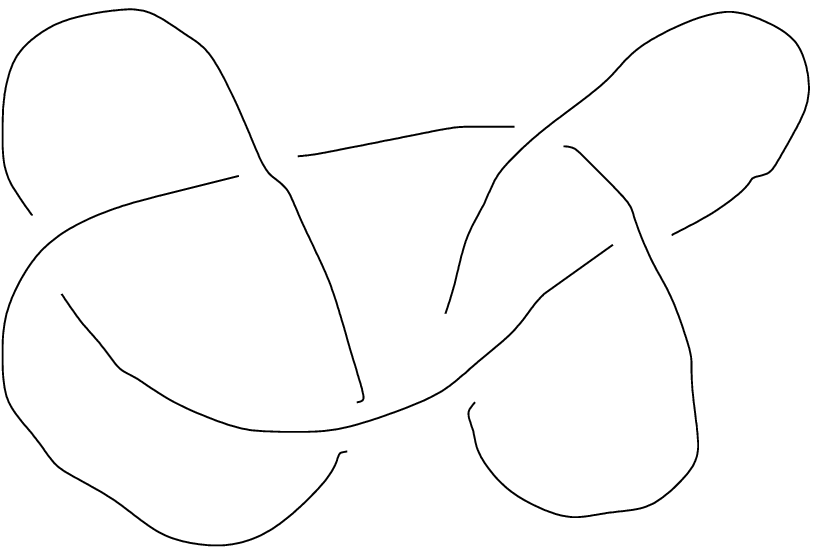} \, \right > \nonumber \\
&=& 3-a^{-4}-a^4+a^{-8}+a^8-a^{-12}-a^{12} \nonumber \\
\label{Knot6}
\end{eqnarray}
since $\omega(L)=0$.

\begin{figure}[h]
\begin{center}
\includegraphics[width=1.9in]{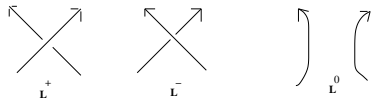}
\end{center}
\caption{Orientation diagrams in the defining relation for $X(L)$.}
\label{Orientation}
\end{figure}

Substituting $a=q^{-1/4}$ into $X(L)$ one obrains $V(L)$ the {\em Jones polynomial} of the oriented link $L$. For the reef knot in eq.(\ref{Knot6}) we have 
\begin{equation}
V\left ( \, \mfig[scale=0.3]{prothi.eps} \, \right )=3-q-q^{-1}+q^2+q^{-2}-q^3-q^{-3}
\label{VKnot6}
\end{equation}

The Jones polynomial satisfies the following relations \cite{VFRJones}
\begin{eqnarray}
q^{-1}V(L^+)-qV(L^{-})&=&(q^{1/2}-q^{-1/2})V(L^0)  
\label{skein1} \\
V(L\sqcup 0)&=&-(q^{-1/2}+q^{1/2})V(L)  \\
\label{skein2}
V(0)&=&1  
\label{skeinr} \\ \nonumber
\end{eqnarray}
Equation (\ref{skein1}) is known as the {\em skein relation}, $L^+, L^-,L^0$ 
are the three link diagrams exhibited in figure \ref{Orientation} .
The condition $L\sqcup 0$ stands for the link $L$ with an added circle 
that does not intersect $L$ (and has no crossing points with $L$). The 
last condition says that the Jones polynomial of the circle is 1.

Using equations (\ref{skein1})-(\ref{skeinr}) one can calculate in an 
alternative way 
the polynomial $V(L)$. Indeed, the skein relation for the figure 8 knot 
reads

\begin{equation}
q^{-1}V\left(\, \mfig[scale=0.3]{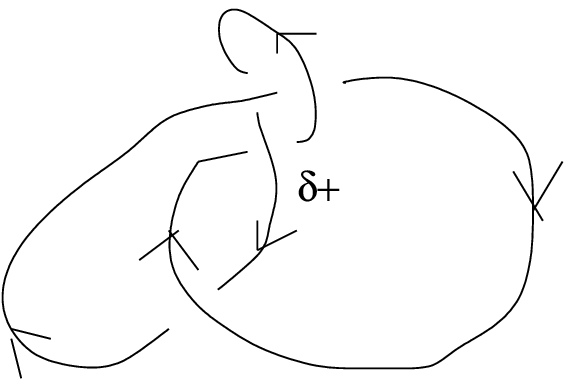} \, \right )-q
V\left(\, \mfig[scale=0.3]{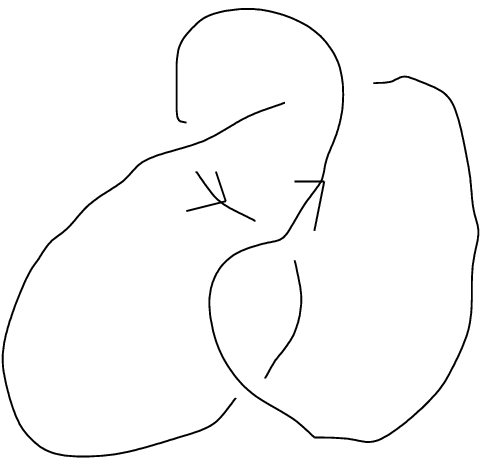} \, \right )=
(q^{1/2}-q^{-1/2})V\left(\, \mfig[scale=0.3]{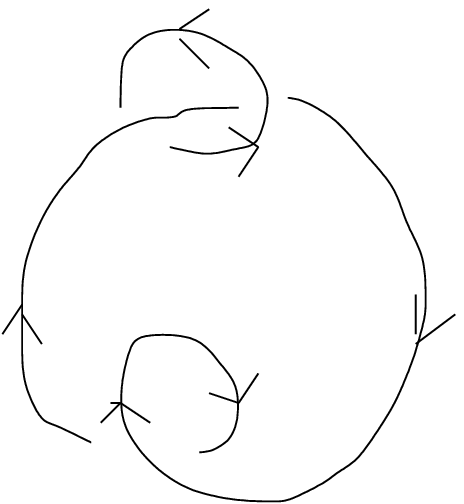} \, \right )
\label{FigureEight}
\end{equation}
or
\begin{equation}
V\left(\, \mfig[scale=0.3]{T4.eps} \, \right )=1+q^2+q^{-2}-q-q^{-1}
\end{equation}
where we used the fact that $V(L)$ for the Hopf link in equation 
(\ref{FigureEight}) is equal to:
$-q^{-1/2}-q^{-5/2}$.
Our result using the skein relations agrees with the previous calculation 
equation (\ref{kn41}),
where the rules for the bracket polynomial have been applied.

Applying the skein relation for the Stevedore's knot once we obtain
\begin{equation}
-q V\left(\,\mfig[scale=0.3]{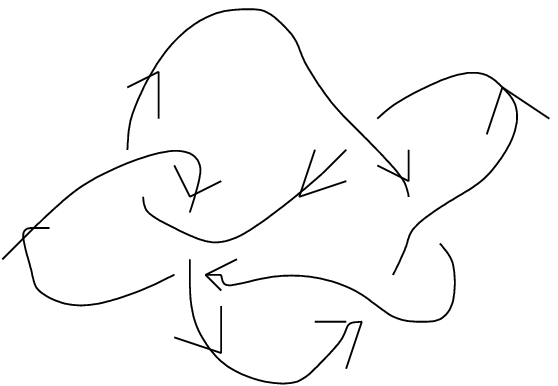} \,\right)+
q^{-1} V\left(\,\mfig[scale=0.3]{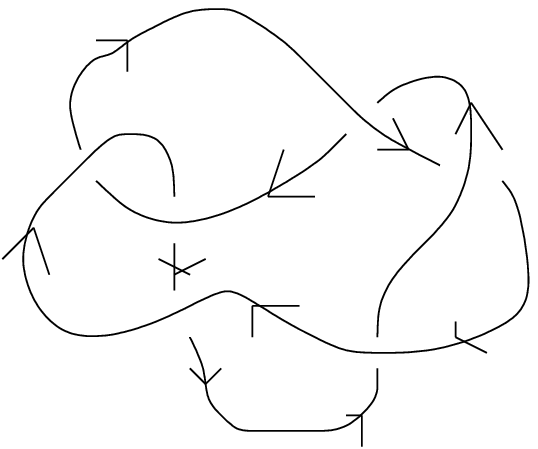}\,\right)=
(q^{1/2}-q^{-1/2})V\left(\,\mfig[scale=0.3]{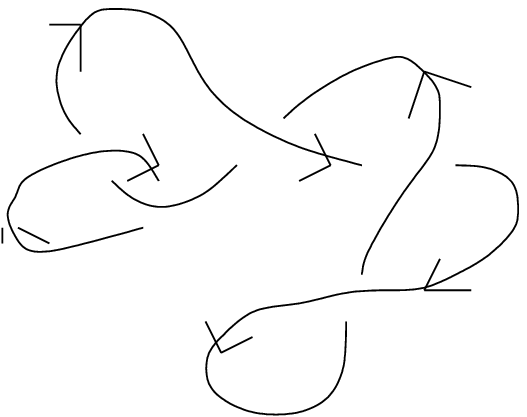}\,\right)
\label{Stevedore}
\end{equation}
The right hand side of equation (\ref{Stevedore}) is a Hopf link 
with Jones polynonial: $-q^{5/2}-q^{1/2}$.
Applying the skein relations to the second member of the left hand side 
of (\ref{Stevedore}) we obtain the unknot and a Hopf link. 
Eventually we obtain for the Jones polynomial of the Stevedore's ($6_1$) 
knot
\begin{equation}
V\left(\,\mfig[scale=0.3]{Stevedore.eps} \,\right)=
2-q-2q^{-1}+q^{-2}+q^2-q^{-3}+q^{-4}
\end{equation}
As a final example the Jones polynomial for the link of the 
Borromean rings is calculated to be:
\begin{equation}
V\left(\,\mfig[scale=0.3]{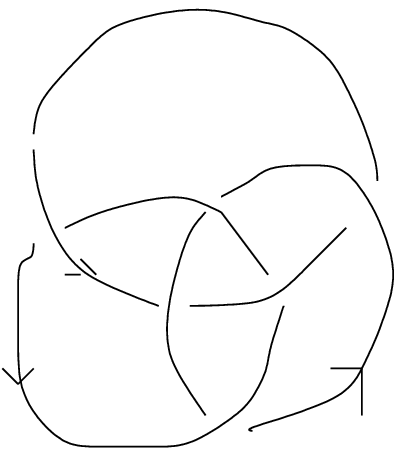} \,\right)=
-q^3+3 q^2-2q+4-2 q^{-1}+3 q^{-2}-q^{-3}
\label{Borromean}
\end{equation}

\subsection{Braid words for links, knots and the HOMFLY polynomial}

It is  useful to express the links and knots we discussed 
in the previous section in words in terms of the generators of the braid 
group. 
The braid word for the Hopf link, in terms of generators of the standard 
and pure braid group is:
$\sigma_1^2=A_{12}$ 
while for the Borromeo rings which appear in Eq. (\ref{Borromean}) the 
corresponding word is given by:
$\sigma_1\sigma_2^{-1}\sigma_1\sigma_2^{-1}\sigma_1\sigma_2^{-1}=A_{12}
A_{13}^{-1}A_{12}^{-1}A_{13}$
On the other hand for the reef or square knot in eq.(\ref{Knot6})
 the braid word is 
$\sigma_1^{-3}\sigma_2^3$, i.e. a braid with 3-strings.

The  two-variable HOMFLY polynomial of the oriented 
link $L$ is defined as follows \cite{FYHLMO},\cite{Jones2}
\begin{equation}
X_L(x,\lambda):=\left(-\frac{1-\lambda x}{\sqrt{\lambda}(1-x)}\right)^{n-1}
(\sqrt{\lambda})^e{\rm tr}(\pi(b))
\end{equation}
where $b\in B_n$ is any braid with $\beta(b)=L$, $e$ being the exponent 
sum of $b$ as a word on the $\sigma_i$'s and $\pi$ the 
representation of $B_n$ in the Hecke algebra $H(x,n)$, $\sigma_i\rightarrow 
g_i$. The Jones one-variable polynomial of the previous section is then 
a special case of the HOMFLY polynomial \cite{VFRJones},\cite{Jones2}
\begin{equation}
V(q)=X_L(q,q)
\end{equation}
For instance for the Hopf link the HOMFLY polynomial is 
calculated to be
\begin{eqnarray}
X_{\rm Hopf\; link}(x,\lambda)&=&\left(-\frac{1-\lambda x}{\sqrt{\lambda}(1-x)}\right)^{2-1}
(\sqrt{\lambda})^2 {\rm tr}(g_1^2) \nonumber \\
&=& -\frac{1-\lambda x}{\sqrt{\lambda}(1-x)}\lambda \;{\rm tr} \left( (x-1)g_1+x \right) \nonumber \\
&=&-\frac{1-\lambda x}{\sqrt{\lambda}(1-x)}\lambda \left((x-1)(-)\frac{1-x}{
1-\lambda x}+x\right) \nonumber \\
&=&-\frac{\sqrt{\lambda}}{1-x}\left(1-x+x^2(1-\lambda)\right)
\end{eqnarray}
and the Jones polynomial $V(q)=X_L(q,q)=-q^{1/2}-q^{5/2}$ which agrees 
with our calculation in the previous section using the skein relations 
or the bracket polynomial.

\section{The Fundamental group of ${\cal D}_n$}
\label{ThemelioOmada1}

The fundamental group of a topological space $X$ can be introduced 
by making the homotopy equivalence classes of paths that start and 
end at a fixed point in a space into a group.
Indeed, for a point $x$ in $X$, a loop at $x$ is a path that starts and 
ends at $x$. Then the {\em fundamental group of X with base point x}, 
denoted $\pi_1(X,x)$, is defined to be the set of equivalence classes 
of loops at $x$, where the equivalence is by homotopy. Also the notions 
of covering spaces and fundamental groups are intimately related:
coverings correspond to subgroups of the fundamental group. There is a 
universal covering, from which all other coverings can be constructed \cite{Fulton}.

In \cite{Terada} the universal covering space $\tilde{\cal D}_n$ of the domain ${\cal D}_n$ 
of Appell-Lauricella's function $F_D$ was determined.
There it was shown that $\tilde{\cal D}_n$ is isomorphic to ${\cal E}$ 
which is the space of the quotient of all simple, closed, rectilinear 
curves on Riemann's sphere by a certain equivalence.
Subsequently, a presentation for $\pi_1({\cal D}_n,a)$ was determined, where 
it was showed  that $\pi_1({\cal D}_n,a)$ is isomorphic to 
$P_{n+2}/Z_{n+2}$. The normal subgroup $Z_{n+2}$ of the pure braid group 
denotes its centre.

We follow the notation used in \cite{Terada}.
Consider the set $N_n=\{0,1,\ldots,n+1\}$ where $n$ is a non-negative integer.
Being given $a,C_a$ on the Riemann's sphere $U$ and two different 
integers $i,j\in N_n$, one considers a simple curve (path):
$u=u_{ij}(t), (0\leq t\leq 1)$ such that 
$u_{ij}(0)=a_i, u_{ij}(1)=a_j,u_{ij}(t)=u_{ji}(1-t)$
and that $u_{ij}(t)$ is contained in the domain $U(C_a)$ provided 
that $0<t<1$ and a family of functions $h_{C_a,ij,s}(t)$ of which the 
curve: $u=h_{C_a,ij,s}(t)$ is a lace in comparison with $u_{ij}(s)$ leaving 
$a_i$. For $I=\{i_{\alpha}\}\subset N_n$, one supposes always 
$i_{\alpha}<i_{\beta}$ if $\alpha<\beta$. 
For each pair $i,j\in N_n$, one denotes by $A_{ij}$ the element of 
$\pi_1({\cal D}_n,a)$ represented by the curve:
$z_{\alpha}=a_{\alpha} (i\not = \alpha), z_i=h_{C_a,ij,1}(t)$ \footnote{As it 
is remarked in 
\cite{Terada} $A_{ij}$ can be regarded as an element of the 
coloured braid group. 
Indeed two sets $(z)$ and $(z^{\prime})$ of $n+2$ complex numbers among whom no two 
can be equal were defined to be equivalent if and only if $z_0-z_0^{\prime}=\cdots
=z_{n+1}-z^{\prime}_{n+1}$. Defining $A_{ij}$ by the curve with the same formulae 
as above in this new space 
one can regard $A_{ij}$ as an element of the pure braid group.
When $i<j$, using the elements $\sigma_{\alpha}(
\alpha=0,\cdots,n)$ of the braid group $B_{n+2}$, one can write 
$A_{ij}=\sigma_i^{-1}\sigma_{i+1}^{-1}\cdots\sigma_{j-2}^{-1}\sigma_{j-1}^2
\sigma_{j-2}\cdots\sigma_i$ and $A_{ji}=\sigma_{j-1}\cdots \sigma_{i+1}\sigma_i^2 
\sigma_{i+1}^{-1}\cdots\sigma_{j-1}^{-1}$.}.
   
Being given a group $G_n$ generated by $\{A_{ij};i,j\in N_n,i \not =j\}$, 
one poses, for $I=\{i_{\alpha};\alpha\in N_p\}$, 
\begin{eqnarray}
& & A_{i_0i_1 \cdots i_p;i_{p+1}}:=A_{i_0i_{p+1}}A_{i_1 i_{p+1}}\cdots A_{i_pi_{p+1}},\nonumber \\
& & A_I=A_{i_0i_1\cdots i_{p+1}}:=A_{i_0;i_1}A_{i_oi_1;i_2}\cdots A_{i_0i_1\cdots i_p;i_{p+1}} \nonumber \\
\label{AnotherGeometry}
\end{eqnarray}

It is said that $G_n$ admits the relation ${\bf R}^n_0$ if one has 
$A_{ij}=A_{ji} \forall i,j\in N_n$ and that $G_n$ admits ${\bf R}_q^n(I)$ if one 
has, for all $J=\{j_{\beta};\beta\in N_q\}$ with $q \leq p$ and $J\subset I, 
A_J \leftrightarrow A_{j_{\alpha}j_{\beta}}$ where $\leftrightarrow$ signifies 
commutativity \cite{Terada}. In addition,  we say  that $G_n$ admits ${\bf R}_n^n$ if one has 
the relation $A_{01\cdots n+1}=1$ the unity of $G_n$ \footnote{Using 
(\ref{AnotherGeometry}) the relation ${\bf R}_n^n$ is equivalent to:
$A_{0;1}A_{01;2}\cdots A_{01\cdots n ; n+1}=A_{01}A_{02}A_{12}\cdots A_{0 n+1}A_{1 n+1}\cdots A_{n n+1}=1$.}.

Being given $I\subset N_n$ and a positive $\eta$ one writes
\begin{eqnarray}
S_I(\eta)&=&\bigcap_{\alpha\in N_p}\{z=(z_0,z_1,\cdots,z_{n+1});z\in{\cal D}_n, \nonumber \\
&&|z_{i_{\alpha}}-z_{i_0}|<\eta \;{\rm sup}\;\{|z_i-z_{i_0}|;i\in N_n\setminus I \}\}
\end{eqnarray}
and \r{S}$_I(\eta)$ is the set of the interior points.

Then using a series of lemmas the author transports the paths and homotopies 
of ${\cal D}_n$ to those of $S_I(\eta)$ and he proves 
that the 
fundamental group $\pi_1({\cal D}_n,a)$ is generated by \cite{Terada}:
$$\{A_{ij};i,j\in N_n,i\not =j\}$$
 and the relations among these elements 
are reduced to the set of relations: 
$${\bf R}_0^n,{\bf R}_1^n(N_n),{\bf R}_2^n(N_n),{\bf R}_n^n.$$ 
Consequently one can chose $(n+1)(n+2)/2-1$ elements as 
generators.

Subsequently the author in \cite{Terada} as a corollary determines that the coloured 
braid group is generated by 
$$\{A_{ij};i,j\in N_n,i\ne j\}$$ 
and the relations 
among the elements are reduced to those of the set of relations 
$${\bf R}_0^n,{\bf R}_1^n(N_n),{\bf R}_2^n(N_n).$$

\section{Conclusions}
\label{Sympera}

In this work using and combining results from \cite{KraniotisKerr}, \cite{KraniotisLight}, \cite{KraniotisSstars} and \cite{Terada} we have established a  
direction connection of the theory of General Relativity with the theory 
of the pure braid group.
More specifically, the connection is established via the 
generalised multivariable hypergeometric function of Lauricella 
$F_D$ through which the exact solutions of the equations  
of motion of test and photon particles in Kerr and Kerr-(anti) de Sitter 
spacetimes and of the corresponding physical quantities such as 
periapsis and gravitomagnetic precessions, bending of light and 
deflection angle were expressed \cite{KraniotisKerr}, \cite{KraniotisLight}, \cite{KraniotisSstars} .
\
As we discussed in the main text the topological properties of the domain 
of variables ${\cal D}_n$ are such that the fundamental group 
$\pi_1({\cal D}_n,a)$ is isomorphic to the quotient group $P_{n+2}/Z_{n+2}$ \cite{Terada}.

The domain of variables ${\cal D}_n$ of $F_D$ is related in the theory 
of General Relativity to the first integrals of motion as well as to the 
cosmological constant and the Kerr (spin) parameter of the rotating 
black hole or rotating central star.

We also mentioned in the main body of the paper that the closure operation 
on pure braids lead to links.

We believe that the link established in this work between General Relativity 
the leading fundamental physical theory of gravity 
and low dimensional topology 
which involves the theory of coloured braids and links is very important.
It may also provide us with hints and clues about the observed dimensionality 
of spacetime and the topological origin of some physical quantities.
On the topology side it might lead to new invariants for links through 
the hypergeometric representation of the pure braid symmetry group.
The first light from the dawn of a new era has reached us.

\section{Acknowledgments}

This work is supported by a Max Planck research fellowship at the 
Max-Planck-Institute for Physics in Munich. 
At early stages  it was partially supported by a fellowship at 
the Ludwig-Maximilians-Universit$\rm\ddot{a}$t in Munich.
The author is grateful to Dieter L$\rm{\ddot u}$st for discussions.

\end{document}